\documentclass[conference,hidelinks]{IEEEtran}

\ifCLASSINFOpdf
\else
\usepackage[dvips]{graphicx}
\fi

\usepackage{amsmath,amssymb,amsfonts}
\usepackage{array}
\usepackage[caption=false,font=normalsize,labelfont=sf,textfont=sf]{subfig}
\usepackage{textcomp}
\usepackage{url}
\usepackage{verbatim}
\usepackage{graphicx}
\usepackage{cite}
\usepackage{xcolor}
\usepackage{orcidlink}
\usepackage[acronym,shortcuts]{glossaries}
\usepackage{bm}
\usepackage{relsize}
\usepackage{soul}
\usepackage{algorithm}
\usepackage{algcompatible}
\usepackage{mathtools}
\usepackage{bm}
\usepackage{lipsum}
\usepackage{mathtools}
\usepackage{dblfloatfix}
\DeclarePairedDelimiter\abs{\lvert}{\rvert}%
\DeclarePairedDelimiter\norm{\lVert}{\rVert}%

\def\BibTeX{{\rm B\kern-.05em{\sc i\kern-.025em b}\kern-.08em
T\kern-.1667em\lower.7ex\hbox{E}\kern-.125emX}}

\newcommand{\trans}[0]{^{\mathsf{T}}}

\newcommand{\herm}[0]{^{\mathsf{H}}}

\newacronym{CSI}{CSI}{channel state information}
\newacronym{LoS}{LoS}{line-of-sight}
\newacronym{NLoS}{NLoS}{non-LoS}
\newacronym{RPE}{RPE}{radar parameter estimation}
\newacronym{OTFS}{OTFS}{orthogonal time frequency space}
\newacronym{AFDM}{AFDM}{affine frequency division multiplexing}
\newacronym{CRLB}{CRLB}{Cram{\`e}r-Rao lower bound}
\newacronym{BCRLB}{BCRLB}{Bayesian Cram{\`e}r-Rao lower bound}
\newacronym{BBI}{BBI}{Bayesian bilinear inference}
\newacronym{AoA}{AoA}{angle-of-arrival}
\newacronym{SNR}{SNR}{signal-to-noise ratio}
\newacronym{ML}{ML}{maximum likelihood}
\newacronym{MIMO}{MIMO}{multiple-input multiple-output}
\newacronym{SISO}{SISO}{single-input single-output}
\newacronym{MUSIC}{MUSIC}{multiple signal classification}
\newacronym{MU}{MU}{multi-user}
\newacronym{ROOT-MUSIC}{ROOT-MUSIC}{ROOT multiple signal classification}
\newacronym{JCAS}{JCAS}{joint communication and sensing}
\newacronym{JCR}{JCR}{joint communications and radar}
\newacronym{ISAC}{ISAC}{integrated sensing and communications}
\newacronym{3D}{3D}{three-dimensional}
\newacronym{2D}{2D}{two-dimensional}
\newacronym{1D}{1D}{one-dimensional}
\newacronym{RX}{RX}{receive}
\newacronym{TX}{TX}{transmit}
\newacronym{BF}{BF}{beamformer}
\newacronym{ROI}{ROI}{region of interest}
\newacronym{mmWave}{mmWave}{millimeter-wave}
\newacronym{MF}{MF}{matched-filter}
\newacronym{DD}{DD}{delay-Doppler}
\newacronym{SotA}{SotA}{state-of-the-art}
\newacronym{ULA}{ULA}{uniform linear array}
\newacronym{QAM}{QAM}{quadrature amplitude modulation}
\newacronym{ISFFT}{ISFFT}{inverse symplectic finite Fourier transform}
\newacronym{SFFT}{SFFT}{symplectic finite Fourier transform}
\newacronym{ISI}{ISI}{inter-symbol interference}
\newacronym{AWGN}{AWGN}{additive white Gaussian noise}
\newacronym{MSE}{MSE}{mean-squared-error}
\newacronym{LMMSE}{LMMSE}{linear minimum mean square error}
\newacronym{RMSE}{RMSE}{root mean square error}
\newacronym{ESPRIT}{ESPRIT}{estimation of signal parameters via rotational invariant techniques}
\newacronym{OFDM}{OFDM}{orthogonal frequency division multiplexing}
\newacronym{OCDM}{OCDM}{orthogonal chirp division multiplexing}
\newacronym{BS}{BS}{base station}
\newacronym{UE}{UE}{user equipment}
\newacronym{JCDE}{JCDE}{joint channel and data estimation}
\newacronym{PDA}{PDA}{probabilistic data association}
\newacronym{PMF}{PMF}{probability mass function}
\newacronym{PBiGaBP}{PBiGaBP}{parametric bilinear Gaussian belief propagation}
\newacronym{PBiGAMP}{PBiGAMP}{parametric bilinear generalized approximate message passing}
\newacronym{GaBP}{GaBP}{Gaussian belief propagation}
\newacronym{FT}{FT}{frequency-time}
\newacronym{DFT}{DFT}{discrete Fourier transform}
\newacronym{IDFT}{IDFT}{inverse discrete Fourier transform}
\newacronym{TD}{TD}{time domain}
\newacronym{wlg}{w.l.g.}{without loss of generality}
\newacronym{CP}{CP}{cyclic prefix}
\newacronym{DAF}{DAF}{discrete affine Fourier}
\newacronym{DAFT}{DAFT}{discrete affine Fourier transform}
\newacronym{IDAFT}{IDAFT}{inverse discrete affine Fourier transform}
\newacronym{CPP}{CPP}{\textit{chirp-periodic} prefix}
\newacronym{IDZT}{IDZT}{inverse discrete Zak transform}
\newacronym{DZT}{DZT}{discrete Zak transform}
\newacronym{P/S}{P/S}{parallel-to-serial}
\newacronym{S/P}{S/P}{serial-to-parallel}
\newacronym{SBL}{SBL}{sparse Bayesian learning}
\newacronym{MPA}{MPA}{message passing algorithms}
\newacronym{EM}{EM}{expectation maximization}
\newacronym{sIC}{soft IC}{soft interference cancellation}
\newacronym{soft RG}{soft RG}{soft replica generation}
\newacronym{BG}{BG}{belief generation}
\newacronym{SGA}{SGA}{scalar Gaussian approximation}
\newacronym{CLT}{CLT}{central limit theorem}
\newacronym{PDF}{PDF}{probability density function}
\newacronym{QPSK}{QPSK}{quadrature phase-shift keying}
\newacronym{ICI}{ICI}{inter-carrier interference}
\newacronym{BER}{BER}{bit error rate}
\newacronym{DoF}{DoF}{degrees-of-freedom}
\newacronym{VGA}{VGA}{vector Gaussian approximation}
\newacronym{FD}{FD}{full-duplex}
\newacronym{SIC}{SIC}{self-interference cancellation}
\newacronym{NMSE}{NMSE}{normalized mean square error}
\newacronym{KL}{KL}{Kullback-Leibler}
\newacronym{LASSO}{LASSO}{least absolute shrinkage and selection operator}
\newacronym{FP}{FP}{fractional programming}
\newacronym{CC}{CC}{communication-centric}

\begin{document}

\title{Blind Bistatic Radar Parameter Estimation for AFDM Systems in Doubly-Dispersive Channels}
\author{\IEEEauthorblockN{Kuranage Roche Rayan Ranasinghe*\textsuperscript{\orcidlink{0000-0002-6834-8877}}, Kengo Ando*\textsuperscript{\orcidlink{0000-0003-0905-2109}}, Hyeon Seok Rou*\textsuperscript{\orcidlink{0000-0003-3483-7629}}, \\ Giuseppe Thadeu Freitas de Abreu*\textsuperscript{\orcidlink{0000-0002-5018-8174}} and Andreas Bathelt$^\dag$\textsuperscript{\orcidlink{0000-0002-5099-5494}}}
\IEEEauthorblockA{\textit{*School of Computer Science and Engineering, Constructor University, Bremen, Germany} \\ \textit{$^\dag$High Frequency Radar and Applications, Fraunhofer FHR, Wachtberg, Germany} \\ 
(kranasinghe,kando,hrou,gabreu)@constructor.university, andreas.bathelt@fhr.fraunhofer.de\\[-3ex]}
}





\maketitle

\begin{abstract}
We propose a novel method for blind bistatic \ac{RPE}, which enables \ac{ISAC} by allowing passive (receive) \acp{BS} to extract radar parameters (ranges and velocities of targets), without requiring knowledge of the information sent by an active (transmit) \ac{BS} to its users.
The contributed method is formulated with basis on the covariance of received signals, and under a generalized doubly-dispersive channel model compatible with most of the waveforms typically considered for \ac{ISAC}, such as \ac{OFDM}, \ac{OTFS} and \ac{AFDM}.
The original non-convex problem, which includes an $\ell_0$-norm regularization term in order to mitigate clutter, is solved not by relaxation to an $\ell_1$-norm, but by introducing an arbitrarily-tight approximation then relaxed via \ac{FP}.
Simulation results show that the performance of the proposed method approaches that of an ideal system with perfect knowledge of the transmit signal covariance with an increasing number of transmit frames.
\end{abstract}


\glsresetall

\IEEEpeerreviewmaketitle
\section{Introduction}
\label{sec:introduction}

Within the recently explored paradigms of enabling technologies for B5G and 6G networks, \ac{ISAC} is one of the fundamental pillars which aims to combine and consolidate the presently separate areas of radar sensing and communications which is expected to significantly improve the performance and efficiency of the system \cite{Wild_Access_2021,Wang_CST_2023,GonzalezProcIEEE2024}.
One of the most commonly addressed \ac{ISAC} approaches is the enabling of sensing functionalities via integration into wireless communications systems \cite{LiyanaarachchiTWC2021,Mohammed_BITS_2022,GaoTWC2023},
referred to as \ac{CC}-\ac{ISAC}.

There are many candidate digital modulation schemes investigated for \ac{CC}-\ac{ISAC} systems \cite{ZhouOJCOMS2022,Rou_SPM_2024}, such as \ac{OFDM} \cite{BarnetoTMTT2019}, \ac{OTFS} \cite{GaudioTWC2020} and \ac{AFDM} \cite{Bemani_WCL_2024} waveforms in which the latter is currently considered to be the most promising, as it was shown to outperform \ac{OFDM} and \ac{OTFS} in high-mobility scenarios in terms of \ac{BER} and
modulation complexity, respectively, while achieving an optimal diversity order in doubly-dispersive channels \cite{BemaniAFDM_ICC_2021, Bemani_TWC_2023}.

Unfortunately, \ac{CC}-\ac{ISAC} systems typically consider the monostatic case of \ac{RPE}, whereby estimation occurs at the actively transmitting \ac{BS}, \textit{i.e.,} at the colocated transmitter and receiver, via the reflected echoes of the transmit signal, which implies: \textit{a)} the need for costly \ac{FD} hardware \cite{KhaledianTMTT2018} or sophisticated \ac{SIC} mechanisms \cite{ZhangBook2023}, and \textit{b)} full knowledge of the transmit signal for \ac{RPE}.

A bistatic \ac{CC}-\ac{ISAC} system, where the transmitter (henceforth referred to as the active \ac{BS}) and receiver (henceforth referred to as the bistatic \ac{BS}) are distributed in space, mitigates the latter problem, but introduces the challenge that the information embedded in the transmit signal is instantaneously unknown at the bistatic \ac{BS} conducting \ac{RPE}.
While the fundamental idea of employing target sensing at a secondary (locally distinct) bistatic receiver resonates with passive radar \cite{KuschelAESM2019, MalanowskiBook2019, GriffithsBook2022}, this technique is not subject to the blindness problem as the reference signal ($i.e.,$ pilot symbols) is deterministically known and exploited.

To circumvent this issue, initial work on bistatic \ac{ISAC} \cite{LeyvaVTC2022} utilizes a cooperative bistatic \ac{BS} topology connected via a fronthaul for \ac{RPE}. 
While the simulation results demonstrate effective \ac{ISAC} performance, the method would in reality be subject to degradation due to imperfection in the fronthaul, which may introduce relative delays and distortion in the signals utilized by the bistatic \ac{BS}.
Another relevant \ac{SotA} work is a novel sensing-aided channel estimation method proposed for \ac{AFDM} in \cite{ZhuWCL2024}, which can be interpreted as a variation of bistatic \ac{ISAC}.
However, this technique still makes use of a preamble ($i.e.,$ pilot symbols) for sensing and the resulting non-blind framework considers a static scattering environment, where only range estimation is performed.

Motivated by the above, we consider an alternative \ul{blind} bistatic \ac{CC}-\ac{ISAC} scenario, in which the bistatic \ac{BS} has \ul{no knowledge} of the information transmitted by the active \ac{BS}.
For such a scenario, we develop our solution under an \ac{AFDM} setting, while offering straightforward generalization to the \ac{OFDM} and \ac{OTFS} waveforms, thanks to the doubly-dispersive channel model utilized \cite{Rou_SPM_2024}.
In particular, we focus on bistatic \ac{RPE}, where a bistatic \ac{BS} receives both the \ac{LoS} signal transmitted by an active \ac{BS}, as well as \ac{NLoS} signals reflected by objects/targets in the surrounding.
As will be shown, the a-priori knowledge of the effective channel structure is actually sufficient for the sensing operation, $i.e.,$ \ac{RPE} can be performed by exploiting the channel structure without requiring the knowledge or seperate estimation of the transmit signal.
Trivially, this bistatic \ac{ISAC} approach requires no \ac{SIC}, and can be considered an enabler of \ac{CC}-\ac{ISAC}.
%

\begin{figure*}[!b]
\vspace{-3ex}
\hrulefill
\setcounter{equation}{2}
\normalsize
\begin{equation}
\label{eq:diagonal_CP_matrix_def}
{
\mathbf{\Phi}_p \triangleq \text{diag}\Big( [ \overbrace{e^{-j2\pi\phi_\mathrm{CP}(\ell_p)}, e^{-j2\pi\phi_\mathrm{CP}(\ell_p-1)}, \dots, e^{-j2\pi\phi_\mathrm{CP}(2)}, e^{-j2\pi\phi_\mathrm{CP}(1)}}^{\ell_p \; \text{terms}}, \overbrace{1, 1, \dots, 1, 1}^{N - \ell_p \; \text{ones}}] \Big) \in \mathbb{C}^{N \times N}.
\vspace{-5ex}}
\end{equation}

%
\setcounter{equation}{3}
\begin{equation}
\label{eq:diagonal_Doppler_matrix_def}
\boldsymbol{\Omega} \triangleq \text{diag}\Big([1,e^{-j2\pi /N},\dots,e^{-j2\pi (N-2) /N}, e^{-j2\pi (N-1) /N}]\Big) \in \mathbb{C}^{N \times N}.
\end{equation}
\setcounter{equation}{0}
\vspace{-3ex}
\end{figure*}

The proposed method leverages multiple frames (transmissions) from an active \ac{BS}, and computes a sample covariance matrix at the bistatic \ac{BS} from the received signals. 
A modified version of this sample covariance matrix, along with a purpose-built dictionary matrix only dependent on the channel grid structure, is then used to formulate a novel sparse recovery problem, which in turn is used to estimate target ranges and velocities via an $\ell_0$-norm regularization leveraging \ac{FP}.
The proposed bistatic \ac{RPE} is performed with a fully blind assmption, with no instantaneous knowledge of the transmit signal at the receive bistatic \ac{BS} or the total number of targets in the surrounding.

The rest of the paper is structured as follows:
\begin{itemize}
\item A concise system model of the doubly-dispersive scattering channel and the signal model of the \ac{AFDM} waveform is provided in Section \ref{sec:system_model}.
\item The proposed blind bistatic \ac{RPE} method including relevant formulations and an original optimization problem is presented in Section \ref{sec:proposed_method}, with the solution of the formulated problem provided in Algorithm \ref{alg:proposed_bistatic}.
\item Numerical simulations and analysis is provided in Section \ref{sec:performance_analysis}, which prove the effectiveness of the proposed blind bistatic \ac{CC}-\ac{ISAC} technique.
\end{itemize}

\vspace{-1ex}
\section{System Model}
\label{sec:system_model}
\vspace{-1ex}

\begin{figure}[H]
    \centering
    \includegraphics[width=0.95\columnwidth]{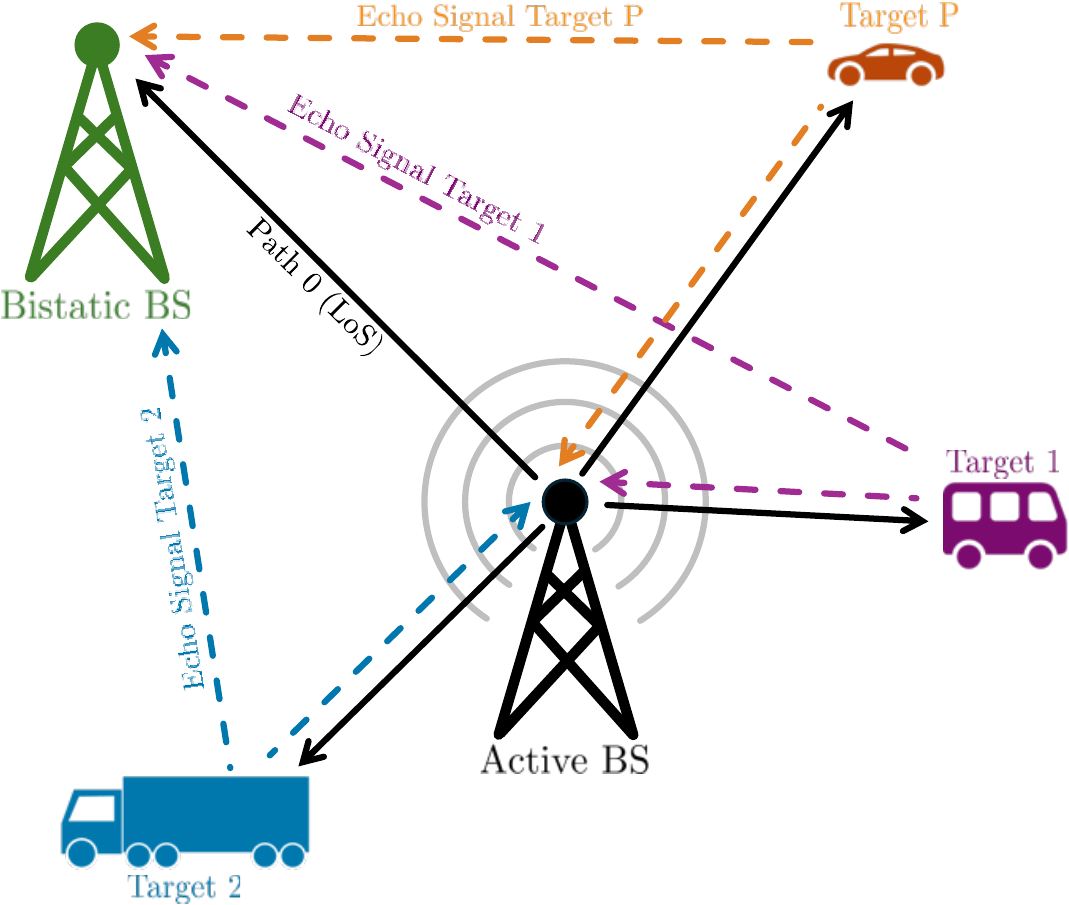}
    \vspace{-1ex}
    \caption{Illustration of an ISAC system showing the 1 \ac{LoS} path and $P$ target echo paths between the active \ac{BS} and the passive bistatic \ac{BS}.}
    \label{fig:ISAC_system_model}
    \vspace{-1ex}
\end{figure}

Consider an \ac{ISAC} scenario composed of one downlink transmitter (active \ac{BS}), one passive receiver (bistatic \ac{BS}) and $P$ significant scatterers in the environment.
As illustrated in Figure \ref{fig:ISAC_system_model} and assuming a point target model \cite{GaudioTWC2020}, there is one \ac{LoS} signal path between the active \ac{BS} and the bistatic \ac{BS}, in addition to the $P$ echo \ac{NLoS} signal paths from each target.
Note that any number of the $P$ scatterers may be communicati-
\vspace{-3ex}
\noindent ons \acp{UE} receiving the downlink signal, but are still considered targets from the bistatic \ac{BS} perspective.

%
%
\vspace{-0.5ex}
\subsection{Doubly-Dispersive Channel Model}
\label{sec:Generalized_Doubly-Dispersive_Channel_Model}
\vspace{-1ex}

We consider the doubly-dispersive wireless channel {\cite{Bliss_Govindasamy_2013,Rou_SPM_2024}} to model the \ac{ISAC} scenario with one \ac{LoS} and $P$ \ac{NLoS} propagation paths, whose channel impulse response function $h(t, \tau)$ in the continuous time-delay domain is described by
\vspace{-1ex}
\begin{equation}
\label{eq:doubly_dispersive_time_delay_channel}
h(t,\tau) \triangleq \sum_{p=0}^P h_p \cdot e^{j2\pi \nu_p t} \cdot \delta(\tau - \tau_p),
\vspace{-0.5ex}
\end{equation}
where $p = 0$ corresponds to the \ac{LoS} path and $p \in \{1,\cdots\!,P\}$ corresponds to the \ac{NLoS} path from each $p$-th target; 
$h_p \!\in\! \mathbb{C}$ is the $p$-th channel fading coefficient;
$\tau_p \!\in\! [0,\!\tau_\text{max}]$ is the $p$-th path delay bounded by the maximum delay $\tau_\text{max}$; 
$\nu_p \!\in\! [-\nu_\text{max},\nu_\text{max}]$ is the $p$-th Doppler shift bounded by the maximum Doppler shift $\nu_\text{max}$
and $\delta(\,\cdot\,)$ is the unit impulse function.

Given an arbitrary time-domain transmit signal $s(t)$, the received signal $r(t) \triangleq  s(t) * h(t,\tau) + w(t)$ where $w(t)$ is the \ac{AWGN}, can be concisely described by a discrete circular convolutional form \cite{Rou_SPM_2024} by sampling at the frequency  $f_\mathrm{S}$, 
\vspace{-1ex}
\begin{equation}
\label{eq:channel_matrix_TD_general}
\mathbf{r} = \!\bigg( \sum_{p=0}^P h_p \!\cdot \mathbf{\Phi}_p 
\!\cdot \boldsymbol{\Omega}^{f_p} \!\cdot\! \mathbf{\Pi}^{\ell_p} \!\bigg) \!\cdot \mathbf{s} + \mathbf{w} = \bm{\Psi} \mathbf{s} + \mathbf{w} \; \in \mathbb{C}^{N \times 1},
\vspace{-1ex}
\end{equation}
where $\mathbf{r}\in \mathbb{C}^{N \times 1}$, $\mathbf{s} \in \mathbb{C}^{N \times 1}$ and $\mathbf{w} \in \mathbb{C}^{N \times 1}$ are the transmit, received and \ac{AWGN} signal vectors consisting of $N$ samples, respectively;
%
%
$\bm{\Psi} \in \mathbb{C}^{N \times N}$ is the effective circular convolutional channel matrix, $\mathbf{\Phi}_p \in \mathbb{C}^{N \times N}$ described in equation \eqref{eq:diagonal_CP_matrix_def} is a diagonal matrix capturing the effect of the \ac{CP} phase with $\phi_\mathrm{CP}(n)$ denoting the waveform-dependent phase function \cite{Rou_SPM_2024} on the sample index $n \in \{0,\cdots,N-1\}$;
$\boldsymbol{\Omega} \in \mathbb{C}^{N \times N}$ described in equation \eqref{eq:diagonal_Doppler_matrix_def} is a diagonal matrix containing the $N$ complex roots of unity;
and $\mathbf{\Pi}\in \{0,1\}^{N \times N}$ is the forward cyclic shift matrix, with elements given by
\vspace{-0.5ex}
\setcounter{equation}{4}
\begin{equation}
\label{eq:PiMatrix}
\pi_{i,j} = \delta_{i,j+1} + \delta_{i,j-(N-1)}\;\; \text{where}\;\; \delta _{ij} \triangleq
\begin{cases}
0 & \text{if }i\neq j,\\
1 & \text{if }i=j.
\end{cases}
\end{equation}

\begin{figure*}[b]
\hrulefill
\setcounter{equation}{9}
\normalsize
\begin{equation}
\label{eq:AFDM_diagonal_CP_matrix_def}
\bm{\varPhi}_p \triangleq \text{diag}\Big( [ \overbrace{e^{-j2\pi c_1 (N^2-2N\ell_p)}, e^{-j2\pi c_1 (N^2-2N(\ell_p-1))}, \cdots, e^{-j2\pi c_1 (N^2-2N)}}^{\ell_p \; \text{terms}}, \overbrace{1, 1, \dots, 1, 1}^{N - \ell_p \; \text{ones}}] \Big) \in \mathbb{C}^{N \times N}.
\vspace{-1ex}
\end{equation}
\setcounter{equation}{5}
\end{figure*}

Furthermore, the roots-of-unity matrix $\boldsymbol{\Omega}$ and the forward cyclic shift matrix $\mathbf{\Pi}$ are respectively exponentiated\footnote{Matrix exponentiation of $\boldsymbol{\Omega}$ is equivalent to an element-wise exponentiation of the diagonal entries, and the matrix exponentiation of $\mathbf{\Pi}^k$ is equivalent to a forward (left) circular shift operation of $k$ indices.} to the power of $f_p \triangleq \frac{N\nu_p}{f_\mathrm{S}}$ and $\ell_p \triangleq \frac{\tau_p}{T_\mathrm{S}}$, which are the normalized digital Doppler frequency and the normalized delay\footnote{It is assumed that the sampling frequency is sufficiently high such that the normalized delays $\ell_p$ are approximated as integers with negligible error, \textit{i.e.,} $\ell_p - \lfloor \frac{\tau_p}{T_\mathrm{S}} \rceil \approx 0$.} of the $p$-th path, respectively, where $T_\mathrm{S} \triangleq \frac{1}{f_\mathrm{S}}$ is the delay resolution.
The two matrix exponentiations capture the effect of the Doppler shift and the integer delay in the circular convolutional channel given in equation \eqref{eq:channel_matrix_TD_general}.

\newpage

\subsection{AFDM Signal Input-Output Relationship}
\label{sec:AFDM_System_Model}

The doubly-dispersive channel model in equation \eqref{eq:channel_matrix_TD_general} is derived for an arbitrary transmitter structure in $s(t)$, such that any waveform such as \ac{OFDM}, \ac{OTFS}, and \ac{AFDM} can be used.
Since a thorough comparison of the waveforms for \ac{ISAC} is provided in \cite{RanasingheARXIV2024}, highlighting the superiority of the \ac{AFDM} for \ac{ISAC} in doubly-dispersive channels, the \ac{AFDM} transmitter structure is elaborated below, and considered in this article henceforth \ac{wlg}.
%

Let $\mathbf{x} \in \mathbb{C}^{N \times 1}$ denote the information vector with elements drawn from an arbitrary complex digital constellation $\mathcal{C}$, with cardinality $Q \triangleq |\mathcal{C}|$ and average symbol energy $\sigma^2_X$.
The corresponding \ac{AFDM} modulated transmit signal of $\mathbf{x}$ is given by its \ac{IDAFT}, $i.e.,$
\begin{equation}
\label{eq:AFDM_moduation}
\mathbf{s}_\text{AFDM} = (\mathbf{\Lambda}_1\herm \mathbf{F}_{N}\herm \mathbf{\Lambda}_2\herm) \cdot \mathbf{x} \in \mathbb{C}^{N \times 1},
\end{equation}
where $\mathbf{F}_N \in \mathbb{C}^{N \times N}$ denotes the $N$-point normalized \ac{DFT} matrix, and the two diagonal chirp matrices $\mathbf{\Lambda}_i$ are defined as
\begin{equation}
\label{eq:lambda_def}
\!\!\!\!\mathbf{\Lambda}_i \!\triangleq\! \mathrm{diag}\Big(\big[1, \cdots\!, e^{-j2\pi c_i n^2}\!\!, \cdots\!, e^{-j2\pi c_i (N-1)^2}\big]\Big) \!\in\! \mathbb{C}^{N \times N}\!,\!\!\!
\end{equation}
where the first central frequency $c_1$ can be selected for optimal robustness to doubly-dispersivity based on the channel statistics \cite{Bemani_TWC_2023}, and the second central frequency $c_2$ can be exploited for waveform design and applications \cite{zhu2023low,rou2024afdm}.

In addition, the \ac{AFDM} modulated signal also requires the insertion of a \ac{CPP} to mitigate the effects of multipath propagation \cite{Bemani_TWC_2023} analogous to the \ac{CP} in \ac{OFDM}, whose multiplicative phase function for equation \eqref{eq:diagonal_CP_matrix_def} is given by $\phi_\mathrm{CPP}(n) = c_1 (N^2 - 2Nn)$ \cite{Rou_SPM_2024}.
Correspondingly, the received \ac{AFDM} signal vector is given by
%
\begin{equation}
\label{eq:AFDM_received vector_in_TD}
\mathbf{r}_\text{AFDM} \triangleq \bm{\Psi} \cdot \mathbf{s}_\text{AFDM} + \mathbf{w} \in \mathbb{C}^{N\times 1}.
\end{equation}

Then, the received signal in equation \eqref{eq:AFDM_received vector_in_TD} is demodulated via the \ac{DAFT} to yield
\begin{align}
\mathbf{y}_\text{AFDM} \!&=\! (\mathbf{\Lambda}_2 \mathbf{F}_{N} \mathbf{\Lambda}_1) \cdot \mathbf{r}_\text{AFDM} \in \mathbb{C}^{N\times 1} \label{eq:AFDM_demodulation}\\
&= \! (\mathbf{\Lambda}_2 \mathbf{F}_{N} \mathbf{\Lambda}_1)\! \cdot \!\bigg(\!\!~\! \sum_{p=0}^P h_p \!\!\cdot\!\bm{\varPhi}_p\! \!\cdot\! \boldsymbol{\Omega}^{f_p} \!\!\cdot\! \mathbf{\Pi}^{\ell_p}\!\!\bigg)\! \cdot \!(\mathbf{\Lambda}_1\herm \mathbf{F}_{N}\herm \mathbf{\Lambda}_2\herm) \!\cdot\! \mathbf{x} \nonumber \\[-0.75ex]
& ~~~+\! (\mathbf{\Lambda}_2 \mathbf{F}_{N} \mathbf{\Lambda}_1)\mathbf{w} \nonumber
, \nonumber
\end{align}
where $\bm{\varPhi}_p$ is the diagonal matrix as described in equation \eqref{eq:AFDM_diagonal_CP_matrix_def}, which inherently incorporates the \ac{AFDM} \ac{CPP} phase function.

In light of the above, the final input-output relationship of \ac{AFDM} over doubly-dispersive channels is given by
\setcounter{equation}{10}
\begin{equation}
\mathbf{y}_\text{AFDM} = \mathbf{G}_\text{AFDM} \cdot \mathbf{x} + \tilde{\mathbf{w}}_\text{AFDM} \in \mathbb{C}^{N\times 1},
\label{eq:DAF_input_output_relation}
\end{equation}
where $\tilde{\mathbf{w}}_\text{AFDM} \triangleq (\mathbf{\Lambda}_2 \mathbf{F}_{N} \mathbf{\Lambda}_1)\mathbf{w} \in \mathbb{C}^{N\times 1}$ is an equivalent \ac{AWGN} vector with the same statistical properties\footnote{This is because the \ac{DAFT} is a unitary transformation \cite{Bemani_TWC_2023}.} as $\mathbf{w}$, and $\mathbf{G}_\text{AFDM} \in \mathbb{C}^{N\times N}$ is the effective \ac{AFDM} channel defined by
\vspace{-1ex}
\begin{align}
\label{eq:DAF_domain_effective_channel}
\!\!\!\!\mathbf{G}_\text{AFDM}\! \triangleq\! \sum_{p=0}^P h_p \!\cdot\! (\mathbf{\Lambda}_2 \mathbf{F}_{N} \mathbf{\Lambda}_1) \!\!\cdot\!\! \big(\bm{\varPhi}_p\! \!\cdot\! \boldsymbol{\Omega}^{f_p} \!\!\cdot\! \mathbf{\Pi}^{\ell_p}\!\big) \!\!\cdot\!\! (\mathbf{\Lambda}_1\herm \mathbf{F}_{N}\herm \mathbf{\Lambda}_2\herm). \!\!\!\!
\end{align}

Due to space limitations, the effective channels and equivalent noise of the \ac{OFDM} and \ac{OTFS} waveforms\footnote{Notice that the \ac{CP} phase matrices $\bm{\Phi}$ are reduced to identity matrices for the \ac{OFDM} and \ac{OTFS} waveforms, as there is no multiplicative \ac{CP} phase \cite{Rou_SPM_2024}.} can be found in equations (12) and (19) of \cite{RanasingheARXIV2024}, respectively.

\section{Blind Bistatic Radar Parameter Estimation}
\label{sec:proposed_method}

In this section, we first build a canonical sparse recovery problem by creating a sample covariance matrix leveraging a reformulation of the system model encapsulating multiple transmission frames with a discretized solution space by considering bounds on the maximum ranges/velocities of the potential targets in the surrounding.
Finally, we offer two solutions for the problem: \textit{a)} A naive \ac{LASSO} formulation that is used to initialize the proposed regularization and \textit{b)} a novel non-convex problem formulation which is solved via the introduction of an arbitrarily-tight approximation relaxed using \ac{FP}.

\subsection{System Model Reformulation}
\label{sec:reformulation}

First, let us express the effective input-output relationship given in equation \eqref{eq:DAF_input_output_relation} for a general waveform as
\vspace{-1ex}
\begin{equation}
\label{eq:generalized_IO_relationship}
\mathbf{y} = \Big(\underbrace{\sum_{p=0}^P h_p \cdot \mathbf{\Gamma}_p}_{\triangleq \mathbf{G}}\Big) \cdot \mathbf{x} + \tilde{\mathbf{w}} \in \mathbb{C}^{N \times 1},
\vspace{-2ex}
\end{equation}
where the matrices $\mathbf{\Gamma}_p$ capture the long-term\footnote{For the purpose of this paper, the \ac{DD} parameters are assumed to stay constant for a sufficient number of transmission frames \cite{RasheedVTC2020}.} \ac{DD} statistics of the channel comprising of the coefficients with the delay and Doppler shifts.
Similarly, the system in equation \eqref{eq:generalized_IO_relationship} without the extrinsic summation on the path index $p$ is given by
\begin{equation}
\label{eq:generalized_IO_relationship_matrix}
\mathbf{y} = \mathbf{H} \cdot \mathbf{\Gamma} \cdot \mathbf{x} + \tilde{\mathbf{w}} \in \mathbb{C}^{N \times 1},
\end{equation}
where the channel coefficient matrix $\mathbf{H}$ and the long-term \ac{DD} matrix $\mathbf{\Gamma}$ are respectively defined as 
\vspace{-0.5ex}
\begin{equation}
\label{eq:matrix_channel_coefs}
\mathbf{H} \! \triangleq \!
\left[
\begin{array}{@{\,}c@{\,}c@{\,}c@{\,}c@{\,}c@{\,}c@{\,}c@{\,}c@{\,}c@{\,}c@{\,}c@{\,}}
h_0 & \cdots & 0 & h_1 & \cdots & 0 & h_P & \cdots & 0\\[-1ex]
\vdots & \ddots & \vdots & \vdots & \ddots & \vdots & \vdots & \ddots & \vdots\\[-0.25ex]
0 & \cdots & h_0 & 0 & \cdots & h_1 & 0 & \cdots & h_P\\
\end{array}
\right]
\in \mathbb{C}^{N \times N(P+1)},
\end{equation}
and
\begin{equation}
\label{eq:concat_matix_Gamma}
\mathbf{\Gamma} \triangleq 
\begin{bmatrix}
\mathbf{\Gamma}_0 & \mathbf{\Gamma}_1 & \cdots & \mathbf{\Gamma}_P
\end{bmatrix}\trans
\in \mathbb{C}^{N(P+1) \times N}.
\vspace{-0.5ex}
\end{equation}

Finally, considering a stream of $T$ transmitted frames, where each frame refers to a single vector $\mathbf{x}\in \mathbb{C}^{N\times 1}$, yields
\begin{equation}
\label{eq:generalized_IO_time_matrix_form}
\mathbf{Y} = \mathbf{H} \cdot \mathbf{\Gamma} \cdot \mathbf{X} + \tilde{\mathbf{W}} \in \mathbb{C}^{N \times T},
\vspace{-1ex}
\end{equation}
where 
\begin{subequations}
\label{eq:sys_model_matrix_components}
\begin{equation}
\mathbf{Y} \triangleq [\mathbf{y}_1, \dots, \mathbf{y}_2, \dots, \mathbf{y}_T] \in \mathbb{C}^{N \times T}, 
\end{equation}
\begin{equation}
\mathbf{X} \triangleq [\mathbf{x}_1, \dots, \mathbf{x}_2, \dots, \mathbf{x}_T] \in \mathbb{C}^{N \times T}, 
\end{equation}
\begin{equation}
\tilde{\mathbf{W}} \triangleq [\tilde{\mathbf{w}}_1, \dots, \tilde{\mathbf{w}}_2, \dots, \tilde{\mathbf{w}}_T] \in \mathbb{C}^{N \times T}. 
\end{equation}
\end{subequations}

\subsection{Proposed Problem Formulation}
\vspace{-0.5ex}

To facilitate a completely blind estimation which implies the availability of only the received signals, we leverage the sample covariance matrix described by
\vspace{-1ex}
\begin{eqnarray}
\label{eq:initial_covariance_formulation}
\mathbf{Y} \cdot \mathbf{Y}\herm = \mathbf{H} \cdot \mathbf{\Gamma} \cdot \mathbf{X} \cdot \mathbf{X}\herm \cdot \mathbf{\Gamma}\herm \cdot \mathbf{H}\herm + \mathbf{H} \cdot \mathbf{\Gamma} \cdot \mathbf{X} \cdot \tilde{\mathbf{W}}\herm  &&
\\ 
&&\hspace{-46.5ex} ~~\;\!+ \tilde{\mathbf{W}} \cdot \mathbf{X}\herm \cdot \mathbf{\Gamma}\herm \cdot \mathbf{H}\herm + \tilde{\mathbf{W}} \cdot \tilde{\mathbf{W}}\herm   \in \mathbb{C}^{N \times N}. \nonumber
\end{eqnarray}
\vspace{-4ex}

Provided that a normalized symbol constellation $\mathcal{C}$ and a sufficiently large $T$ is used, equation \eqref{eq:initial_covariance_formulation} can be rewritten as\footnote{Since $\tilde{\mathbf{W}}$ is matrix constructed from $T$ \ac{AWGN} vectors, a trivial computation yields the result $\tilde{\mathbf{W}} \cdot \tilde{\mathbf{W}}\herm = \sigma^2_W \cdot T \cdot \mathbf{I}_N$.}
\vspace{-2ex}
\begin{equation}
\label{eq:cov_matrix}
\mathbf{Y} \!\cdot\! \mathbf{Y}\herm \approx T (\sigma^2_X \cdot \mathbf{H} \cdot \mathbf{\Gamma} \cdot \mathbf{\Gamma}\herm \cdot \mathbf{H}\herm + \sigma^2_W \cdot \mathbf{I}_N)  \in \mathbb{C}^{N \times N},
\vspace{-1ex}
\end{equation}
where $\sigma^2_W$ is the variance of an arbitrary column in $\tilde{\mathbf{W}}$.

Setting $\sigma^2_X = 1$ \ac{wlg}, equation \eqref{eq:initial_covariance_formulation} can also be written as
\vspace{-1.5ex}
\begin{equation}
\label{eq:modified_cov_matrix}
\!\!\!\!\underset{T \rightarrow \infty}{\text{lim}} \underbrace{\frac{1}{T} \!\cdot\! \mathbf{Y} \!\cdot\! \mathbf{Y}\herm \!-\! \sigma^2_W \!\cdot\! \mathbf{I}_N}_{\triangleq \tilde{\mathbf{R}}_Y \in \mathbb{C}^{N \times N}} \!= \!\underbrace{\mathbf{H} \cdot \mathbf{\Gamma} \cdot \mathbf{\Gamma}\herm \cdot \mathbf{H}\herm}_{\text{perfect covariance matrix}} \!\in \mathbb{C}^{N \times N},
\vspace{-1ex}
\end{equation}
where we have implicitly defined the modified sample covariance matrix $\tilde{\mathbf{R}}_Y \in \mathbb{C}^{N \times N}$.

Next, using the definition for $\mathbf{H} \cdot \mathbf{\Gamma}$ and equation \eqref{eq:generalized_IO_relationship}, equation \eqref{eq:modified_cov_matrix} can be expressed in terms of two sums as
\vspace{-1ex}
\begin{equation}
\label{eq:cov_matrix_w_sums}
\tilde{\mathbf{R}}_Y = \bigg(\sum_{p=0}^P h_p \cdot \mathbf{\Gamma}_p\bigg) \!\cdot\! \bigg(\sum_{p=0}^P h_p \cdot \mathbf{\Gamma}_p\bigg)^{\!\!\mathsf{H}} \in  \mathbb{C}^{N \times N}.
\vspace{-1ex}
\end{equation}

Defining $\mathbf{H}_\mathrm{G} \triangleq \mathbf{h} \cdot \mathbf{h}\herm$ with $\mathbf{h} \triangleq [h_0, \dots, h_p, \dots, h_P]\trans$, the $(p,q)$-th element can be represented by $\mathbf{H}_\mathrm{G}(p,q)$.
Using this definition, simplifying equation \eqref{eq:cov_matrix_w_sums} yields
\vspace{-1ex}
\begin{equation}
\label{eq:cov_matrix_expanded_sums}
\tilde{\mathbf{R}}_Y = \sum_{p=0}^P \sum_{q=0}^P \mathbf{H}_\mathrm{G}(p,q) \cdot \underbrace{\mathbf{\Gamma}_p \cdot \mathbf{\Gamma}_q\herm}_{\tilde{\mathbf{\Gamma}}_{p,q}} \in  \mathbb{C}^{N \times N}.
\vspace{-2ex}
\end{equation}

Since each distinct matrix $\mathbf{\Gamma}_p$ is unitary, the matrices $\tilde{\mathbf{\Gamma}}_{p,q}$ in equation \eqref{eq:cov_matrix_expanded_sums} amount to identity matrices of size $N\times N$ when $p\!=\!q$, resulting in uniqueness ($i.e.,$ useful distinct information for \ac{RPE}) only for $p \! \neq \! q$.
Since this uniqueness is preserved in mostly diagonal/off-diagonal elements in these sparse matrices and to alleviate the computational burden of using $N\times N$ matrices, we utilize a summation of the rows of equation \eqref{eq:cov_matrix_expanded_sums}
\vspace{-3ex}
\begin{align}
\label{eq:cov_matrix_expanded_sums_summingop}
\bar{\mathbf{r}}_Y \;\triangleq \tilde{\mathbf{R}}_Y \!\cdot\! \mathbf{1}_N = \! \sum_{p=0}^{P} \sum_{q=0}^{P} \mathbf{H}_\mathrm{G}(p,q) \cdot \!\!\! \underbrace{\tilde{\mathbf{\Gamma}}_{p,q} \!\cdot\! \mathbf{1}_N}_{\triangleq \mathbf{e}_{p,q} \in \mathbb{C}^{N \times 1}} \!\!\!= \mathbf{E} \!\cdot\! \text{vec}(\mathbf{H}_\mathrm{G}), \nonumber \\[-3.5ex]
\end{align}
\vspace{-4ex}

\noindent where $\mathbf{1}_N$ denotes an $N \times 1$ vector of ones, $\text{vec}(\mathbf{H}_\mathrm{G})$ denotes the column-wise vectorized form of $\mathbf{H}_\mathrm{G}$, and the dictionary matrix $\mathbf{E} \in \mathbb{C}^{N \times (P+1)^2}$ is defined as
\vspace{-1ex}
\begin{equation}
\label{eq:sparse_dict_matrix_def}
\mathbf{E} \triangleq \big[\mathbf{e}_{0,0}, \cdots\!, \mathbf{e}_{0,P}, \cdots\!, \mathbf{e}_{P,0}, \cdots\!, \mathbf{e}_{P,P}\big] \in \mathbb{C}^{N \times (P+1)^2}.
\vspace{-1ex}
\end{equation}

Next, we follow a similar strategy in \cite{Mehrotra_TCom_2023,RanasingheARXIV2024} to sparsify the system by decoupling and discretizing the path-wise doubly-dispersive channel model, \textit{i.e.,} reformulating the delay-Doppler representation of equation \eqref{eq:doubly_dispersive_time_delay_channel} given by \cite{Rou_SPM_2024} 
\vspace{-1.5ex}
\begin{equation}
\label{eq:DD_channel_definition}
h(\tau,\nu) = \sum_{p=0}^P h_p \!\cdot\! \delta(\tau - \tau_p) \!\cdot\! \delta(\nu - \nu_p),
\vspace{-1ex}
\end{equation}
into
\vspace{-1ex}
\begin{equation}
\label{eq:DD_channel_definition_frac_Dopp}
h(\tau,\nu) = \sum_{k=0}^{K_\tau - 1} \sum_{d=0}^{D_\nu - 1} h_{k,d} \!\cdot\! \delta(\tau - \tau_k) \!\cdot\! \delta(\nu - \nu_d),
\vspace{-1ex}
\end{equation}
where $K_\tau \triangleq \mathrm{max}(\ell_p)$ and $D_\nu \triangleq \mathrm{max}(f_p)$ are respectively the maximum normalized delay spread and maximum digital normalized Doppler spread indices satisfying the underspread assumption $K_\tau <\!< N$ and $D_\nu <\!< N$.

Consequently, the normalized delay and Doppler shift parameters are decoupled from each $p$-th path and instead into a sparsified delay-Doppler grid\footnote{We remark that if the spacing of such a grid is sufficiently set, the only non-zero $h_{k,d}$ in equation \eqref{eq:DD_channel_definition_frac_Dopp} are those in which both $\tau_k\approx \tau_p$ and $\nu_d \approx \nu_p$, which can be exploited to perform \ac{RPE} amounting to the estimation of the $(P+1)$ channel gains such that $h_{k,d}\neq 0$.
This induces a higher intrinsic sparsity since most elements of the vector to be estimated are zeros, paving the path for the use of compressive sensing schemes \cite{Mehrotra_TCom_2023}.}, such that the auxiliary channel matrix is instead described by $\bar{\mathbf{H}}_\mathrm{G} \triangleq \bar{\mathbf{h}} \cdot \bar{\mathbf{h}}\herm$ with $\bar{\mathbf{h}} \triangleq [h_0, \cdots\!, h_{k \cdot d}, \cdots\!, h_{(K_\tau-1)\cdot(D_\nu-1)}]\trans$, consequently yielding
\vspace{-1ex}
\begin{equation}
\label{eq:cov_matrix_sparse_can_form}
\bar{\mathbf{r}}_Y = \bar{\mathbf{E}} \cdot \text{vec}(\bar{\mathbf{H}}_\mathrm{G})  \in  \mathbb{C}^{N \times 1},
\vspace{-1ex}
\end{equation}
with the modified dictionary matrix $\bar{\mathbf{E}} \in \mathbb{C}^{N \times K_\tau^2 D_\nu^2}$ adopting distinct values for the delay and Doppler indices given by
\vspace{-1ex}
\begin{eqnarray}
\label{eq:sparse_dict_matrix_def_DD}
\bar{\mathbf{E}} \triangleq \big[\bar{\mathbf{e}}_{0,0}, \cdots\!, \bar{\mathbf{e}}_{0,(K_\tau-1)\cdot(D_\nu-1)}, \cdots &&
\\ 
&& \hspace{-34ex} \cdots\!, \bar{\mathbf{e}}_{(K_\tau-1)\cdot(D_\nu-1),0}, \cdots\!, \bar{\mathbf{e}}_{(K_\tau-1)\cdot(D_\nu-1),(K_\tau-1)\cdot(D_\nu-1)}\big], \nonumber
\end{eqnarray}
\vspace{-3ex}

\noindent with $\bar{\mathbf{e}}_{k\cdot d,k' \cdot d'} \in \mathbb{C}^{N \times 1}$ correspondingly defined as
\vspace{-1ex}
\begin{align}
\bar{\mathbf{e}}_{k \cdot d,k' \cdot d'} \triangleq &(\mathbf{\Lambda}_2 \mathbf{F}_{N} \mathbf{\Lambda}_1) \cdot (\bar{\bm{\varPhi}}_k \cdot \bar{\boldsymbol{\Omega}}^{f_d}\cdot \bar{\mathbf{\Pi}}^{\ell_k})\nonumber\\
&\cdot (\bar{\bm{\varPhi}}_{k'} \cdot \bar{\boldsymbol{\Omega}}^{f_{d'}}\cdot \bar{\mathbf{\Pi}}^{\ell_{k'}})\herm \cdot (\mathbf{\Lambda}_1\herm \mathbf{F}_{N}\herm \mathbf{\Lambda}_2\herm)\cdot\mathbf{1}_N,
\end{align}
\vspace{-3.5ex}

\noindent where $\bar{\bm{\varPhi}}_{k}$, $\bar{\boldsymbol{\Omega}}^{f_{d}}$, and $\bar{\mathbf{\Pi}}^{\ell_{k}}$ are computed via equations \eqref{eq:AFDM_diagonal_CP_matrix_def}, \eqref{eq:diagonal_Doppler_matrix_def} and  \eqref{eq:PiMatrix}, respectively, with the indices $\ell_p$ and $f_p$ replaced by their corresponding location on the discrete grid as $\ell_k$ and $f_d$ for each corresponding normalized delay and Doppler index of the two corresponding grid points $(k,d)$ and $(k',d')$.

\vspace{-1ex}
\subsection{Proposed Solution via the $\ell_0$-norm Approximation}
\vspace{-0.5ex}

Observing equation \eqref{eq:cov_matrix_sparse_can_form}, it is most likely that $N <\!< K_\tau^2 D_\nu^2$ for moderate grid resolutions and $\text{vec}(\bar{\mathbf{H}}_\mathrm{G})$ is sparse, such that the estimation of $\bar{\mathbf{H}}_\mathrm{G}$ becomes an underdetermined linear problem, resulting in an infeasible solution via a simple least squares formulation.
Therefore, relaxed approaches such as the \ac{LASSO} \cite{TroppTIT2006} can be leveraged to exploit the intrinsic sparsity of $\bar{\mathbf{H}}_\mathrm{G}$ and obtain a feasible solution via the optimization problem described by
\vspace{-1.5ex}
\begin{equation} 
\label{eq:LASSO_problem}
\underset{\bar{\mathbf{H}}_\mathrm{G}\in\mathbb{C}^{K_\tau D_\nu \times K_\tau D_\nu } }{\text{minimize}}  \norm{ \bar{\mathbf{r}}_Y - \bar{\mathbf{E}} \cdot \text{vec}(\bar{\mathbf{H}}_\mathrm{G}) }^2_2 \,+\, \beta \cdot \norm{\text{vec}(\bar{\mathbf{H}}_\mathrm{G})}_1,
\vspace{-1ex}
\end{equation}
where $\norm{\cdot}_q$ denotes the $\ell_q$-norm, and $\beta \in \mathbb{R}^+$ is a sparsity-enforcing penalty parameter.

However, to complement the full structure of the problem in equation \eqref{eq:cov_matrix_sparse_can_form}, the complete optimization to solve it without the $\ell_1$-norm relaxation of the \ac{LASSO} can be formulated as
\vspace{-1.5ex}
\begin{subequations}
\label{eq:oracle_form}
\begin{align}
\underset{\bar{\mathbf{H}}_\mathrm{G}\in\mathbb{C}^{K_\tau D_\nu \times K_\tau D_\nu } }{\text{minimize}}&  \norm{ \bar{\mathbf{r}}_Y - \bar{\mathbf{E}} \cdot \text{vec}(\bar{\mathbf{H}}_\mathrm{G}) }^2_2,\\
\text{subject to} \hspace{3ex}& \bar{\mathbf{H}}_\mathrm{G} \succcurlyeq 0, \\
&\norm{\text{vec}(\bar{\mathbf{H}}_\mathrm{G})}_0 = (P+1)^2,\\
&\hspace{0.2ex} \mathrm{rank}(\bar{\mathbf{H}}_\mathrm{G})=1,
\end{align}
\end{subequations}
\vspace{-4ex}

\noindent where the first constraint intrinsically forces constructively coupled solutions\footnote{Since $\bar{\mathbf{H}}_\mathrm{G} \triangleq \bar{\mathbf{h}} \cdot \bar{\mathbf{h}}\herm$ by definition, non-zero entries in $\bar{\mathbf{h}}$ will reflect as symmetric coupled entries on the upper and lower triangluar sections of $\bar{\mathbf{H}}_\mathrm{G}$ leading to the positive semidefinite constraint in equation \eqref{eq:oracle_form}.},
the second constraint enforces the sparsity of the solution, and the third constraint trivially results from the definition $\bar{\mathbf{H}}_\mathrm{G} \triangleq \bar{\mathbf{h}} \cdot \bar{\mathbf{h}}\herm$.

\newpage

As a fully blind sensing scenario cannot assume prior knowledge of the actual number of paths in the estimation procedure, in addition to the non-convex third constraint\footnote{While relevant relaxation methods exist in literature \cite{CaoEUSIPCO2017}, this constraint is completely dropped for simplicity and is relegated to a follow-up journal version of this article.}, equation \eqref{eq:oracle_form} is relaxed as
\vspace{-1.5ex}
\begin{subequations}
\label{eq:propose_opt}
\begin{align}
\hspace{-2ex}\underset{\bar{\mathbf{H}}_\mathrm{G}\in\mathbb{C}^{K_\tau D_\nu \times K_\tau D_\nu } }{\text{minimize}}&  \norm{ \bar{\mathbf{r}}_Y \!-\! \bar{\mathbf{E}} \cdot\text{vec}(\bar{\mathbf{H}}_\mathrm{G})}^2_2 + \eta \norm{\text{vec}(\bar{\mathbf{H}}_\mathrm{G})}_0,\!\!\!\!\! \\
\text{subject to} \hspace{3ex}& \bar{\mathbf{H}}_\mathrm{G} \succcurlyeq 0, 
\end{align}
\end{subequations}
\vspace{-4ex}

\noindent where $\eta\in\mathbb{R}^+$ denotes the weight parameter of the $\ell_0$-norm regularization.

The above optimization problem in equation \eqref{eq:propose_opt} is still non-convex due to the $\ell_0$-norm within the objective function, and is consequently convexized via the method proposed in \cite{IimoriOJCOM2021}, which approximates the $\ell_0$-norm of an arbitrary complex vector $\mathbf{b}=[b_1,\dots,b_N]\trans \in\mathbb{C}^{N\times 1}$ as
\vspace{-1ex}
\begin{equation}
\label{eq:l_0_norm_approx}
\norm{\mathbf{b}}_0 \approx \sum_{n=1}^N \frac{\abs{b_n}^2}{\abs{b_n}^2 + \alpha} = N - \sum_{N=1}^M \frac{\alpha}{\abs{b_n}^2 + \alpha},
\vspace{-1ex}
\end{equation}
with $\alpha \in \mathbb{R}^+$ denoting the hyperparameter that controls the tightness of the approximation. 

Applying \ac{FP} \cite{ShenTSP2018} to the $\ell_0$-norm approximation in equation \eqref{eq:l_0_norm_approx} to convexize the affine-over-convex ratios, we obtain
\vspace{-1ex}
\begin{equation}
\label{eq:fract_program_approx}
\norm{\mathbf{b}}_0 \approx N - \Bigg( \sum_{n=1}^N 2 \tilde{\alpha}_n \sqrt{\alpha} - \tilde{\alpha}_n^2 \cdot \abs{b_n}^2 + \tilde{\alpha}_n^2 \cdot \alpha  \Bigg),
\vspace{-1ex}
\end{equation}
where $\tilde{\alpha}_n$ is an auxiliary variable computed from a previously estimated $\tilde{\mathbf{b}}=[\tilde{b}_1,\dots,\tilde{b}_N]\trans$, via
\vspace{-1ex}
\begin{equation}
\label{eq:FP_aux_variable}
\tilde{\alpha}_n \triangleq \frac{\sqrt{\alpha}}{\abs{\tilde{b}_n}^2 + \alpha},\;\forall n .
\end{equation}
\vspace{-2ex}

Consequently, using the above iterative approximation of the $\ell_0$-norm, the optimization problem in equation \eqref{eq:propose_opt} can be relaxed, resulting in
\vspace{-1ex}
\begin{subequations}
\label{eq:propose_opt_convex}
\begin{align}
\underset{\bar{\mathbf{H}}_\mathrm{G}\in\mathbb{C}^{K_\tau D_\nu \times K_\tau D_\nu}\hspace{-3ex} }{\text{minimize}}& \hspace{2.7ex} \norm{ \bar{\mathbf{r}}_Y \!-\! \bar{\mathbf{E}} \cdot\text{vec}(\bar{\mathbf{H}}_\mathrm{G}) }^2_2
\\ & \hspace{3.5ex} + \eta \cdot\text{vec}(\bar{\mathbf{H}}_\mathrm{G})\herm \cdot \tilde{\mathbf{A}} \cdot\text{vec}(\bar{\mathbf{H}}_\mathrm{G}), \nonumber \\
\text{subject to} \hspace{1.7ex}& \hspace{3ex} \bar{\mathbf{H}}_\mathrm{G} \succcurlyeq 0, 
\end{align}
\end{subequations}
where $\tilde{\mathbf{A}} \triangleq \text{diag}([\tilde{\alpha}_1^2, \dots, \tilde{\alpha}_m^2, \dots, \tilde{\alpha}_M^2])$.

\vspace{-1.5ex}
\begin{algorithm}[H]
    \caption{Blind Bistatic Radar Parameter Estimation}
    \label{alg:proposed_bistatic}
    \textbf{Input:} Received signal matrix $\mathbf{Y}$, dictionary matrix $\bar{\mathbf{E}}$, noise power $\sigma_w^2$, number of transmit frames $T$, maximum number of iterations $i_\text{max}$, optimization hyperparameters $\beta$, $\eta$ and $\alpha$. \\
    \textbf{Output:} Estimated delay and Doppler shifts $\hat{\tau}_p$ and $\hat{\nu}_p$. 
    \hrule
    
    \begin{algorithmic}[1]
    \vspace{1ex}
    \STATE Compute $\bar{\mathbf{r}}_Y$ using eqs. \eqref{eq:modified_cov_matrix} and \eqref{eq:cov_matrix_expanded_sums_summingop}.
    \STATE Obtain an initial estimate for $\bar{\mathbf{H}}_\mathrm{G}$ by solving the \ac{LASSO} problem in eq. \eqref{eq:LASSO_problem}.
    \STATEx \hspace{-1ex}{\textbf{for}} {$i=1$ to $i_\text{max}$} {\textbf{do}}
    \STATE Compute the \ac{FP} auxiliary matrix $\tilde{\mathbf{A}}$ via eq. \eqref{eq:FP_aux_variable}.
    \STATE Obtain $\bar{\mathbf{H}}_\mathrm{G}$ by solving the final problem in eq. \eqref{eq:propose_opt_convex}. 
    \STATEx \hspace{-1ex}{\textbf{end}} {\textbf{for}}
    \STATE Compute the estimates $\hat{\tau}_p$ and $\hat{\nu}_p$ corresponding to the $(P+1)^2$ non-zero entries of $\bar{\mathbf{H}}_\mathrm{G}$ in accordance to the dictionary matrix given in eq. \eqref{eq:sparse_dict_matrix_def_DD}; $i.e.,$ match the computed indices to the dictionary matrix to find the final estimates for $\hat{\tau}_p$ and $\hat{\nu}_p$.
    \end{algorithmic}
\end{algorithm}
\vspace{-6ex}

In summary, using the received signal matrix $\mathbf{Y}$ to compute the modified sample covariance vector according to equations \eqref{eq:modified_cov_matrix} and \eqref{eq:cov_matrix_expanded_sums_summingop} and constructing the modified dictionary matrix using equation \eqref{eq:sparse_dict_matrix_def_DD} with all the possible values of $\ell_k$ and $f_d$, we first obtain an initial solution via the \ac{LASSO} formulation in equation \eqref{eq:LASSO_problem} to initialize the final optimization problem.
Next, the final optimization problem in equation \eqref{eq:propose_opt_convex} is iteratively solved to obtain a final estimate for $\bar{\mathbf{H}}_\mathrm{G}$.

Finally, since the dictionary matrix built via equation \eqref{eq:sparse_dict_matrix_def_DD} is constructed using all the possible values of $\ell_k$ and $f_d$, by associating the non-zero entries in $\bar{\mathbf{H}}_\mathrm{G}$ to the columns of $\bar{\mathbf{E}}$, the final estimates for $\hat{\tau}_p$ and $\hat{\nu}_p$ can be obtained.

An overview of the proposed scheme is provided in the form of a pseudo-code in Algorithm \ref{alg:proposed_bistatic}.

\vspace{-1ex}
\section{Performance Analysis}
\label{sec:performance_analysis}
\vspace{-1ex}

To evaluate the performance of the proposed blind bistatic \ac{RPE} scheme, we first consider the illustration detailed in Figure \ref{fig:ISAC_system_model}.
In such a scenario with a passive bistatic \ac{BS} receiving \ac{AFDM}-modulated signals from an active \ac{BS}, we consider blind \ac{RPE} at the bistatic \ac{BS}.

The chosen performance metric is the classical \ac{RMSE} defined as $\Upsilon \triangleq ||\hat{\vartheta}_p - \vartheta_p||^2_2$, where $\vartheta_p$ denotes a given radar parameter and $\hat{\vartheta}_p$ its estimate.

In addition, the \ac{SNR} for \ac{RPE} is defined similarly to \cite{Bemani_WCL_2024} as
\vspace{-1ex}
\vspace{-0.5ex}
\begin{equation}
\label{eq:SNR_def}
\text{SNR} \triangleq \frac{(P+1) \cdot \sigma_h^2}{\sigma_W^2},
\vspace{-1ex}
\end{equation}
where $\sigma_h^2$ denotes the power of the channel coefficient $h_p$ associated with a given $p$-th target, as defined in equation \eqref{eq:doubly_dispersive_time_delay_channel}, already incorporating the \ac{LoS} path.

\begin{figure}[H]
    \centering
    \captionsetup[subfloat]{labelfont=small,textfont=small}
    \subfloat[Range RMSE.]{\includegraphics[width=0.95\columnwidth]{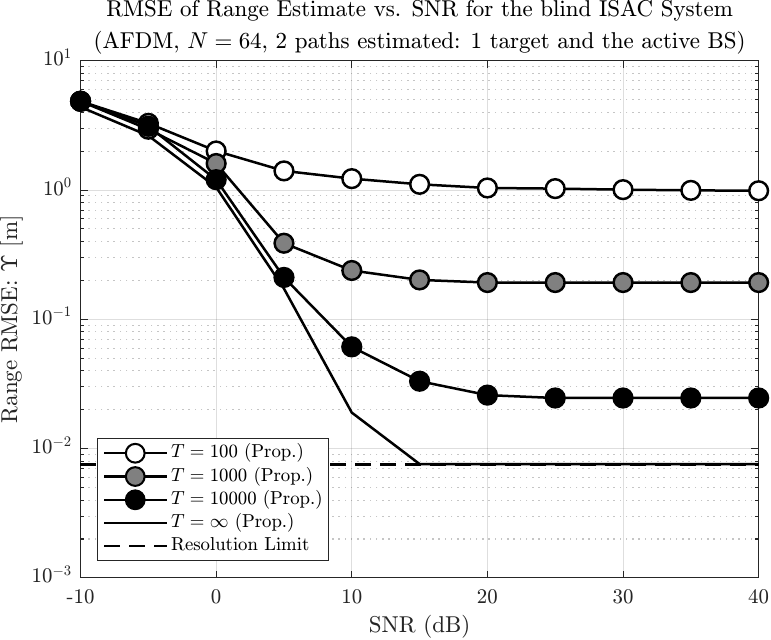}}%
    \vspace{-1.5ex}
    \label{fig:range_plot}
    \subfloat[Velocity RMSE.]{\includegraphics[width=0.95\columnwidth]{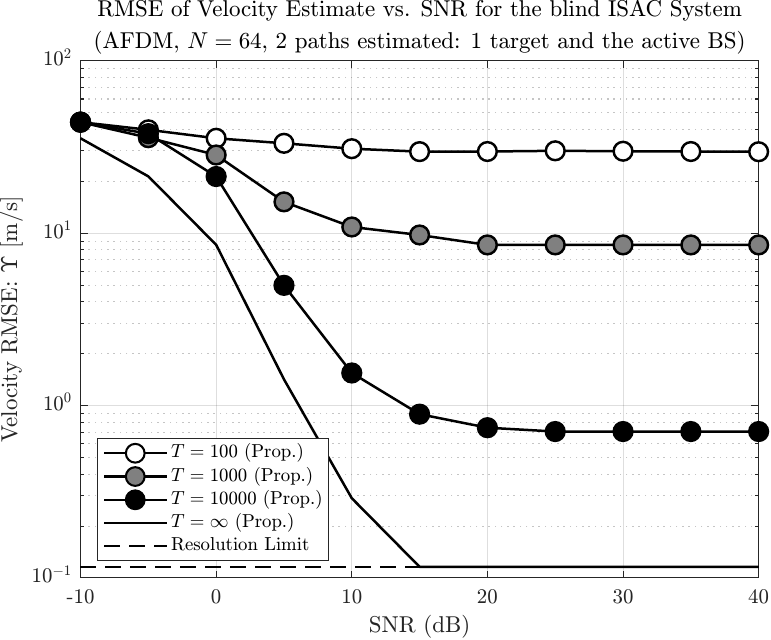}}%
    \label{fig:vel_plot}
    \vspace{-0.5ex}
    \caption{\ac{RMSE} Performance versus \ac{SNR} of the proposed blind bistatic \ac{RPE} scheme over the \ac{AFDM} waveform with a single \ac{LoS} signal from the active (static) \ac{BS} and one \ac{NLoS} echo signal from a target in the surrounding.}
    \label{fig:XXXM_Performance_plot}
\end{figure}

We show, in Figure \ref{fig:range_plot} and Figure \ref{fig:vel_plot}, results for a situation with one \ac{LoS} signal from the transmitting active \ac{BS} which is assumed to be static and situated at a distance of $3.75$ m from the bistatic \ac{BS} and one \ac{NLoS} reflected signal from a target ($i.e., P=1$) at a range of $15$ m moving with a velocity of $37$ m/s towards the bistatic \ac{BS}.
We remark here that since a \ac{LoS} path always exists in the surrounding defining the active \ac{BS}, there will always be more that one target to be detected by the estimation algorithm, which concurrently fits with the positive semidefinite constraint in use in equation \eqref{eq:propose_opt_convex}.

The remaining system parameters for the results shown in Figures \ref{fig:range_plot} and \ref{fig:vel_plot} are as follows: $70 \, \text{GHz}$ central carrier frequency, with a bandwidth of $20 \, \text{MHz}$, \ac{QAM} modulation and maximum normalized delay and digital Doppler shift indices $\ell_\text{max} = 4$ and $f_\text{max} = 0.1$, respectively.
In order to reduce complexity, we also set $N = 64$ for \ac{AFDM}.
Consequently, the proposed blind \ac{RPE} algorithm employs the optimization hyperparameters $\beta = 1$, $\eta = 0.1$, $\alpha = 0.001$ and is run up to $i_\text{max} = 3$ iterations.

As seen from Figures \ref{fig:range_plot} and \ref{fig:vel_plot}, the \ac{RMSE} performance increases significantly when an increasing amount of received frames are utilized during the construction of the sample covariance matrix, with the case for an infinite number of transmit frames converging with the resolution limit \cite{Ranasinghe_ICASSP_2024} obtained when the true $\bar{\mathbf{H}}_\mathrm{G}$ is estimated under \ac{AFDM} waveforms.

\newpage

\vspace{-3ex}
\section{Conclusion}
\label{sec:conclusion}
\vspace{-1ex}

In this paper, we contributed a new \ac{CC}-\ac{ISAC} scheme, namely, a blind bistatic \ac{RPE} technique for any arbitrary communications waveform. 
This was achieved by formulating a novel covariance-based problem, then solved completely by introducing an arbitrarily-tight approximation for the $\ell_0$-norm term in the optimization problem via \ac{FP}.
Finally, we proved the efficacy of the the proposed method with computer simulations, which demonstrated the performance by the way of the \ac{RMSE} for both range and velocity.

\vspace{-1ex}
\bibliographystyle{IEEEtran}

\end{document}